\documentclass[12pt]{article}

\usepackage{hyperref}
\usepackage[left=2cm,right=2cm, top=2cm,bottom=3cm]{geometry}
\usepackage{amsmath}
\usepackage{amsfonts}
\usepackage{amssymb}
\usepackage{graphicx,pgf}
\usepackage{multimedia}
\usepackage{caption}
\usepackage{color}
\usepackage{longtable}
\usepackage{hhline}
\usepackage{xcolor}
\usepackage{indentfirst}

\begin{document}

\title{Some Features of the Modified Solid-Liquid-Vapor Equation of State}

\author{
	Alexey V. Batov \textit{batovalex@yandex.ru},\\
	Ivan A. Galyaev \textit{ivan.galyaev@yandex.ru},\\
	Maksim I. Kostiuchek \textit{max31@list.ru},\\
	Anton M. Salnikov \textit{anton.m.salnikov@hotmail.com}\\\\
	V.A. Trapeznikov Institute of Control Sciences, Russian Academy of Sciences\\
	65 Profsoyuznaya street, Moscow, 117997, Russia
}

\maketitle

\begin{abstract}
    The paper considers the geometry of the Modified Solid-Liquid-Vapour equation of state. This model describes a substance state in three phases. Thermodynamics states are points on Legendrian or Lagrangian manifolds in the corresponding contact or symplectic spaces in terms of differential geometry. The conditions of applicable states and the first order phase transition are given for the Modified Solid-Liquid-Vapour equation of state. The Lagrangian manifold, singularity curve and the phase transition curves are plotted for methane.
\end{abstract}


\section{Introduction}
\label{sec:intro}

Differential geometry is a strong tool for a thermodynamics description. Gibbs was one of the first authors to propose a geometrical representation of thermodynamic properties using surfaces~\cite{bib:Gibbs_1873}. The author studied the thermodynamic surfaces and their properties for solid, liquid, and vapor bodies. Moreover, Gibbs considered critical and triple points. Later Mrugala~\cite{bib:Mrugala_1978} generalized that of Gibbs given. Mrugala adapted a contact manifold as a basic mathematical structure. The empirical laws of thermodynamics have been reformulated using this manifold and exterior differential forms. Arnold made a valuable contribution to the development of the differential geometry approach to thermodynamics. He developed symplectic geometry in particular theory of Lagrangian and Legendrian manifolds which is a mathematical basis for modern thermodynamics. For example see~\cite{bib:Arnold_1985}. The differential geometry approach is continuing to expand nowadays. Lychagin~\cite{bib:Lychagin_2022} introduced a Riemannian structure on the Lagrangian and Legendrian manifolds as well as higher-order symmetric differential forms, introduced for the description of higher-order phase transitions.

The Lagrangian and Legendrian manifolds are defined by two equations of state (EoS) and a thermodynamic potential. Choosing the equations of state is defined by a choice of coordinates on the Lagrangian and Legendrian manifolds. Usually, the thermic and caloric equations of states are used together with the entropy. The volume, the temperature, and the entropy serve as coordinates on manifolds. The thermic EoS represents pressure as a function of the volume and the temperature, while the caloric EoS expresses the internal energy in terms of the volume and the temperature. Moreover, these coordinates on the manifolds define the thermodynamic potential, from which other potentials and equations of state are derived~\cite{bib:Lychagin_2022}.

Many classical thermic equations of state like van der Waals (vdW) equation, Redlich-Kwong equation \cite{bib:Redlich_1949}, Peng-Robinson equation (PR)~\cite{bib:Peng_1976}, have the form
\begin{equation*}
	p = p_{\mathrm{HC}} + p_{\mathrm{A}},
\end{equation*}
where~$p_{\mathrm{HC}}$ is the repulsive term and $p_{\mathrm{A}}$ is the attractive term. Usually, those equations of state describe liquid and vapor phases and only one phase transition. A $p$-$v$ diagram of these equations do not have discontinuity and may have one critical point.

There are modifications of those equations, which describe three phases: liquid, vapor, and solid. Modification of the vdW equation of state has been proposed by Yokozeki in the work~\cite{bib:Yokozeki_2003}. This one was named Solid-Liquid-Vapor equation of state (SLV EoS). Yokozeki has suggested an algorithm for calculating the parameters of the SLV EoS and illustrated results for argon. The SLV EoS has the discontinuity of the isotherms in the $p$-$v$ state diagram. This discontinuity separates the solid phase from the liquid and vapor phases. Additionally, the solid-liquid and solid-vapor phase transition curves have the discontinuity and have no the critical point.

A similar equation with a modified attractive term has been considered in works~\cite{bib:Stringari_2014}.

In the work~\cite{bib:Mo_2022} another similar equation of state has been investigated. This equation has been named the Modified Solid-Liquid-Vapor equation of state (MSLV EoS). An algorithm for calculating the EoS parameters has been suggested. The results have been shown for some matters, in particular methane and ethane. The MSLV EoS has the following properties:
\begin{enumerate}
	\item
	The predicted gas- and liquid-phase properties should be consistent with the PR equation, and the phase change between the gas and liquid phases should be continuous. There is a critical point of the gas-liquid phase change that should be consistent with the form of the ideal gas equation of state.

	\item
	 There should be a stable solid phase in a particular region where the gas and liquid phases exist in the phase diagram. The pressure change of the solid phase with volume should be similar to that of the liquid phase; considering the actual physical significance, there is no critical point of phase change between the solid and liquid phases.
	 
	 \item
	  The prediction of thermodynamic properties, such as the volume and density, of the solid phase at different pressures and temperatures can be consistent with the changing pattern of the actual substance; the solid-liquid-gas properties that change with the pressure and temperature can be described.
\end{enumerate}

In this paper, we apply differential geometry to describe the thermodynamics set by the Modified Solid-Liquid-Vapor equation of state~\cite{bib:Mo_2022}. The results of our study are shown for methane.

The paper has the following structure. The~\ref{sec:intro}st section gives the actuality of the work. In the~\ref{sec:therm}nd we remind the thermodynamics and construction that we will use in the next. The~\ref{sec:EoS_Disc}rd section contains a description of the MSLV EoS. The~\ref{sec:calor}th gives the caloric EoS and potentials. In the~\ref{sec:app_stt}th and~\ref{sec:ph_tr}th we discuss the applicable states and phase transition for the MSLV EoS. The~\ref{sec:sim}th contains numerical results for methane. The~\ref{sec:concl}th gives the conclusion. Additionally, the paper has an appendix where other graphical results for numerical computation are introduced.

\section{Thermodynamics}
\label{sec:therm}

Description of real gas thermodynamics in terms of differential geometry has been suggested in~\cite{bib:Lychagin_2022}. It was shown that the thermodynamic states of real gas are the Legendrian or the Lagrangian manifolds in the corresponding contact or symplectic spaces. These spaces are equipped with a quadratic differential form. The applicable state domain of the thermodynamic model is the one, where this form determines a Riemannian structure. This fact allows us to find the applicable state domain and to find singularities of the Lagrangian manifold projections to spaces of intensive and extensive variables. It is exactly a submanifold, where the quadratic differential form changes its type. Projection singularities, at which intensive variables are preserved and extensive changes abruptly, correspond to phase transitions.

Moreover, symplectic (and contact) geometry and the corresponding Poisson and Lagrange brackets are used to obtain the state equations that define the Lagrangian manifold. This is done by introducing the Massieu-Plank potential.

Consider a differential 1-form in space~$\mathbb{R}^7 (S, E, V, N, T, p, \gamma)$
\begin{equation*}
    \tilde{\theta} = dS - T^{-1} dE - T^{-1} p dV + T^{-1} \gamma dN
\end{equation*}
where~$S$ is the entropy, $E$ is the internal energy, $V$ is the volume, $N$ is the amount of substance, $T$ is the temperature, $p$ is the pressure, $\gamma$ is the chemical potential. Pair~$(\mathbb{R}^7,\,\theta)$ forms contact space. Then the first and second laws of thermodynamics take the form~$\theta = 0$, i.e. the possible thermodynamic states form the Legendre manifold~$\tilde{L} \subset \mathbb{R}^7$ and~$\tilde{\theta}|_{\tilde{L}} = 0$. According to the Gibbs–Duhem principle the Legendrian manifold~$\tilde{L}$ is invariant to the one-parameter group of a contact transformation of extensive quantities:
\begin{equation*}
    (S, E, V, N, T, p, \gamma) \rightarrow (tS, tE, tV, tN, T, p, \gamma).
\end{equation*}
This requirement allows to reduce the dimension of contact space and Legendrian manifold. Let is move on to molar quantities
\begin{equation*}
    \sigma = \frac{S}{N}, \quad
    \varepsilon = \frac{E}{N}, \quad
    V = \frac{V}{N}.
\end{equation*}
In this terms we have
\begin{equation*}
    \tilde{\theta} = N (d\sigma - T^{-1} d\varepsilon - T^{-1} p dv) +
    (\sigma - T^{-1} \varepsilon - T^{-1} p v + T^{-1} \gamma) dN.
\end{equation*}
Thus we have the Legendrian manifold~$\hat{L}$ in contact space~$\mathbb{R}^5 (\sigma, \varepsilon, v, T, p)$ with the contact form
\begin{equation*}
    \theta = d\sigma - T^{-1} d\varepsilon - T^{-1} p dv
\end{equation*}
and an expression for the chemical potential as the molar Gibbs free energy
\begin{equation*}
    \gamma = \varepsilon + p v - T \sigma.
\end{equation*}

We can exclude entropy by a projection of contact space~$\mathbb{R}^5 (\sigma, \varepsilon, v, T, p)$ to symplectic space~$\mathbb{R}^4 (\varepsilon, v, T, p)$ with natural symplectic form
\begin{equation*}
    -d\theta = d(T^{-1}) \wedge d\varepsilon + d(T^{-1} p) \wedge dv.
\end{equation*}
The image of the Legendrian manifold~$\hat{L}$ under the projection is the Lagrangian manifold~$L \subset \mathbb{R}^5$. Let~$v$ and~$T$ be coordinates on~$L$. Then the manifold~$L$ is defined by two equations of state:
\begin{equation*}
    p = a(v,\,T), \quad
    \varepsilon = b(v,\,T),
\end{equation*}
where~$a = a(v,\,T)$ and~$b = b(v\,T)$ are some functions. From the condition of Lagrangian we have
\begin{equation*}
    \left( \frac{a}{T} \right)_T = \left( \frac{b}{T^2} \right)_v,
\end{equation*}
therefore exist a function~$\phi = \phi(v,\,T)$ such that
\begin{equation*}
    \frac{a}{T} = R\phi_v, \quad
    \frac{b}{T^2} = R\phi_T,
\end{equation*}
where~$R$ is the universal gas constant. The function~$\phi$ is called the Massieu-Plank potential. Thus the equations of state take the form
\begin{itemize}
    \item thermic state equation~$p = R T \phi_{v}$;
    \item caloric state equation~$\varepsilon = R T^2 \phi_T$,
\end{itemize}

Other thermodynamic potentials in therm of the Massieu-Plank potential:
\begin{enumerate}
    \item Molar entropy
    \begin{equation*}
        \sigma = R(\phi + T\phi_T);
    \end{equation*}
    \item Molar enthalpy
    \begin{equation*}
        \eta = \varepsilon + p v = R T (T\phi_T + v \phi_{v});
    \end{equation*}
    \item Molar Gibbs free energy
    \begin{equation*}
        \gamma = \varepsilon + p v - T\sigma = R T (v \phi_{v} + \phi).
    \end{equation*}
\end{enumerate}

The applicable states are defined by the differential form
\begin{equation*}
    \kappa = d T^{-1} \cdot d \varepsilon + d \left( p T^{-1} \right) \cdot dv,
\end{equation*}
that must be negative in the applicable states on the Lagrangian manifold~$L$. Where~$\cdot$ denotes the symmetric product of tensors. In therm of the Massieu-Plank potential, the differential form~$\kappa$ take the form
\begin{equation*}
    \label{eq:kappa}
    \kappa = \phi_{vv} dv \cdot dv - \frac{T\phi_{TT} + 2\phi_T}{T} dT \cdot dT.
\end{equation*}
The condition of negative differential form~$\kappa$ gives inequalities for the applicable states
\begin{equation}
    \label{eq:AppSt}
    \phi_{vv} < 0, \quad
    T\phi_{TT} + 2\phi_T > 0.
\end{equation}
or in a more familiar form
\begin{equation*}
    p_v < 0, \quad
    \varepsilon_T > 0.
\end{equation*}

Note that the differential 2-form~(\ref{eq:kappa}) define the Riemannian structure on the Lagrangian manifolds~$L$ in the applicable state domain.

Projection singularities of the Lagrangian manifold on the plane of intensive quantities~$(T,p)$ and plane of extensive quantities~$(v,\varepsilon)$ are defined by expressions~$\phi_{vv} = 0$ and $T\phi_{TT} + 2\phi_T = 0$ respectively. These curves will be named singular curves.

The critical point is called a point that is the solution of the equations
\begin{equation*}
    p_{v} = 0, \quad
    p_{vv} = 0.
\end{equation*}
Or, in therm of the Massieu-Plank potential
\begin{equation*}
    \phi_{vv} = 0, \quad
    \phi_{vvv} = 0,
\end{equation*}

Conditions for preserving of the intensive quantities while changing the extensive quantities give the first order phase transition. Namely the pressure, temperature and molar Gibbs free energy preserved at the molar volume change. Thereby we have
\begin{gather*}
    \phi_{v} (v_2,T) - \phi_{v} (v_1,T) = 0 \\
    \phi (v_2,T) - \phi (v_1,T) - v_2\phi_{v} (v_2,T) + v_1\phi_{v} (v_1,T) = 0,
\end{gather*}
where~$v_1$, $v_2$ denote molar volume and~$v_1 \neq v_2$.

\section{Equation of state}
\label{sec:EoS_Disc}

\begin{figure}[h!!!]
    \centering
    \includegraphics[width=100mm]{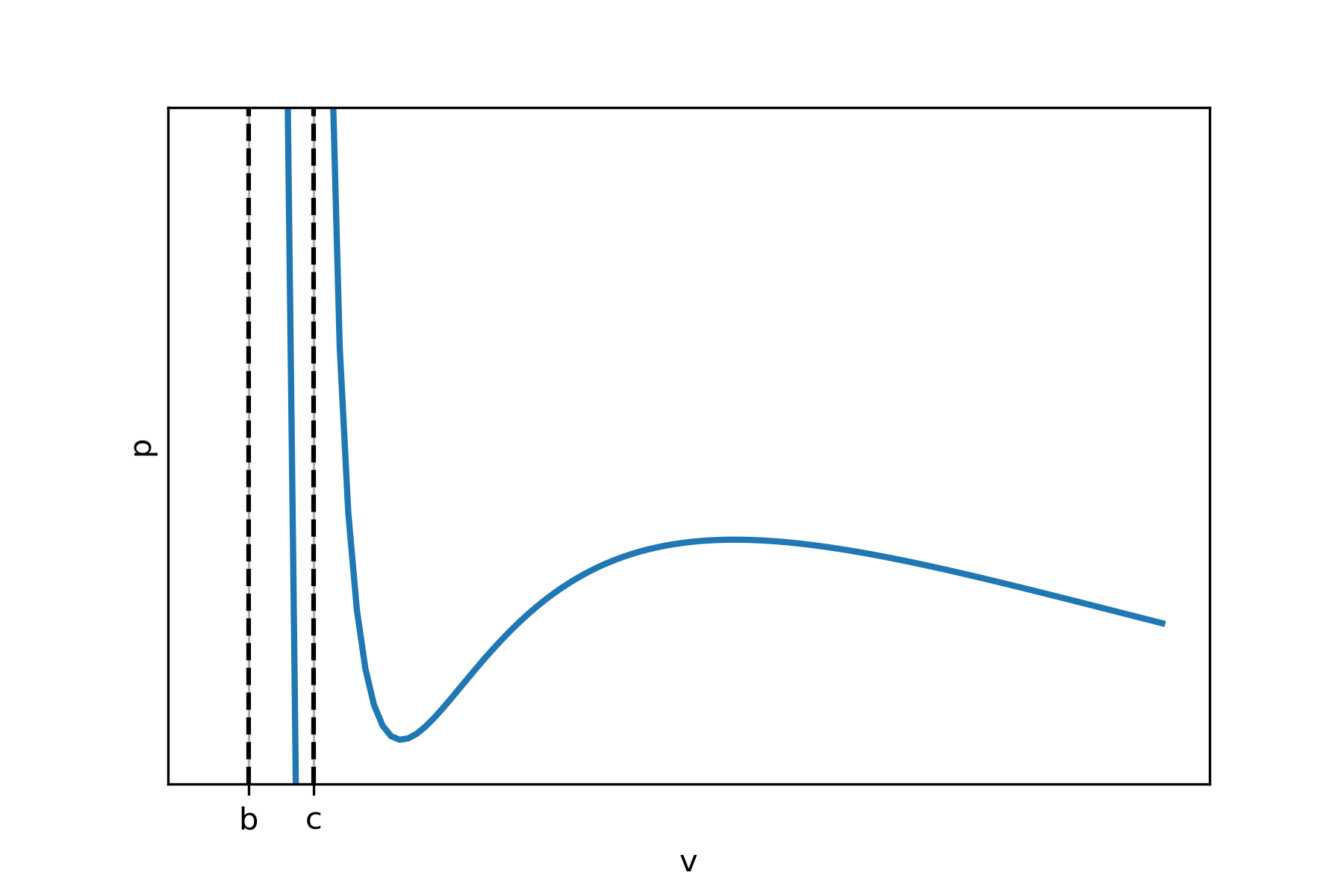}
    \caption{Schematic isotherm in $(v,p)$ at T less then the critical temperature}
    \label{fig:isotherm}
\end{figure}

The Modified Solid-Liquid-Vapor equation of state~\cite{bib:Mo_2022}
\begin{equation}
    \label{eq:EoS1}
    p = \frac{R T}{v-b} \frac{v-d}{v-c} - \frac{q(T)}{v^2 + 2 b v - b^2}, \quad
    T > 0, \quad b < v < d \cup v > c,
\end{equation}
where $p$ is the pressure (MPa); $T$ is the temperature (K); $v$ is the molar volume ($\text{cm}^3/\text{mol}$), $R = 8.314463$ is the universal gas constant ($\text{J} \text{K}^{-1} \text{mol}^{-1}$);
\begin{gather*}
    q(T) = a \alpha(T), \quad
    \alpha(T) = \left( 1 + m \left( 1-\sqrt{\frac{T}{T_c}} \right) \right)^2; \\
    m =
    \begin{cases}
	0.37464 + 1.54226\omega - 0.26992\omega^2, & \omega < 0.491 \\
	0.374642 + 1.48504\omega - 0.164423\omega^2 + 0.016666\omega^3, & \omega > 0.491
    \end{cases};
\end{gather*}
where $a$, $b$, $c$, $d$, $\omega$~are the constants various for each gas, herewith~$b < d < c$; $T_c$~is the critical temperature. It is assumed that~$b$ is the minimum molar volume of the solid, $d$~is the maximum molar volume of the solid and~$c$ is the minimum volume of the liquid. The schematic isotherm of the MSLV EoS is shown in fig.~\ref{fig:isotherm}.

The equation~(\ref{eq:EoS1}) is discontinuous at~$v = d$ and $v = c$. Therefore the Lagrangian manifold is divided into two lists: only for the solid phase at~$b < v < c$ and for both liquid and vapor phases at~$v > c$.

Denote
\begin{equation*}
    f(v) = \frac{R}{v-b} \frac{v-d}{v-c}, \quad
    g(v) = -\frac{1}{v^2 + 2 b v - b^2}.
\end{equation*}
Then the equation of state~(\ref{eq:EoS1}) will take the form
\begin{equation*}
    p = T f(v) + q(T) g(v).
\end{equation*}

\section{The caloric equation of state and thermodynamics potentials}\label{sec:calor}

The Massieu-Plank potential is obtained from the thermic EoS~(\ref{eq:EoS1})
\begin{equation*}
    \phi_v = \frac{f(v)}{R} + \frac{q(T)}{R T} g(v).
\end{equation*}
and
\begin{equation*}
    \phi = \frac{F(v)}{R} + \frac{q(T)}{R T} G(v) + K(T),
\end{equation*}
where
\begin{gather*}
    F(v) = \int f(v) dv = \frac{R}{b-c} \big( (b-d) \ln(v-b) - (d-c) \ln(v-c) \big) + \operatorname{const}, \\
    G(v) = \int g(v) dv = \frac{1}{\sqrt{2} b} \operatorname{artanh} \left( \frac{v+b}{\sqrt{2} b} \right) + \operatorname{const}.
\end{gather*}
and $K(T)$ is some function of~$T$.

Thus the caloric EoS take the form
\begin{equation*}
    \varepsilon = \big( T q'(T) - q(T) \big) G(v) + T^2 K'(T),
\end{equation*}
where~$\varepsilon$ is molar internal energy.

If~$a=b=c=d=0$, the thermic EoS~\eqref{eq:EoS1} is the ideal gas EoS, and it is known that equation for the internal energy for ideal gas has the form~$\varepsilon = \frac{n}{2} R T$, where $n$ is the number degree of freedom. From here we suppose that~$R T^2 K'(T) = \frac{n}{2} R T$, $K(T) = \frac{n}{2} \ln T$.

Thus the Massieu-Plank potential for the MSLV EoS take the form
\begin{equation*}
    \phi = \frac{F(v)}{R} + \frac{q(T)}{R T} G(v) + \frac{n}{2} \ln T,
\end{equation*}
and caloric equation of state take the form
\begin{equation*}
    \varepsilon = \big( T q'(T) - q(T) \big) G(v) + \frac{n}{2} R T.
\end{equation*}

We have the following expressions for the thermodynamics potentials of the MSLV EoS:
\begin{gather*}
    \sigma = F(v) + q'(T) G(v) + \frac{n}{2} R (\ln T + 1), \\
    \eta = \big( T f(v) + q(T) g(v) \big) v + \big( T q'(T) - q(T) \big) G(v) + \frac{n}{2} R T \\
    \gamma = T(v f(v) - F(v)) + q(T) (v g(v) - G(v)) + \frac{n}{2} R T \ln T,
\end{gather*}
where~$\sigma$ is the molar entropy, $\eta$ is the molar enthalpy and $\gamma$ is the molar Gibbs free energy.

\section{Applicable states}
\label{sec:app_stt}

The equation of singularity curve for the MSLV EoS takes the form
\begin{gather}
    \notag
    \phi_{vv} = \frac{f'(v)}{R} + \frac{q(T)}{R T} g'(v) = 0, \\
    \label{eq:SingC2}
    T\phi_{TT} + 2\phi_T = \frac{q''(T) G(v)}{R} + \frac{n}{2 T} = 0,
\end{gather}
where
\begin{equation*}
    q''(T) = \frac{a}{2 T^{3/2} \sqrt{T_c}} (m^2 + m).
\end{equation*}

The function~$T$ versus~$v$ on singularities curve~$\phi_{vv} = 0$ can be found by change of variable~$t=\sqrt{T}$. Thus we obtain
\begin{gather}
    \label{eq:tphivv1}
    t_1 = \frac{(m+1) \sqrt{a g'(v)}}{m \sqrt{a g'(v)} + \sqrt{-f'(v)}}, \\
    \label{eq:tphivv2}
    t_2 = \frac{(m+1) \sqrt{a g'(v)}}{m \sqrt{a g'(v)} - \sqrt{-f'(v)}}
\end{gather}

\section{Phase transition}
\label{sec:ph_tr}

The conditions of phase transition for the MSLV equation of state take the form
\begin{gather}
    \label{eq:PhTr1}
    f(v_2) - f(v_1) + \frac{q(T)}{T}(g(v_2) - g(v_1)) = 0, \\
    \label{eq:PhTr2}
    F(v_2) - F(v_1) - v_2 f(v_2) + v_1 f(v_1) + \frac{q(T)}{T} (G(v_2) - G(v_1) - v_2 g(v_2) + v_1 g(v_1)) = 0,
\end{gather}
where~$v_1$, $v_2$ is the molar volume at the phase transition.

We can find the expression for~$T$ like function of~$v_1$, $v_2$ at the phase transition. The equation
\begin{equation*}
    q(\tau) - \tau^2 h = 0.
\end{equation*}
is obtained from the equation~\eqref{eq:PhTr1}, where
\begin{equation*}
    h = -\frac{f(v_2)-f(v_1)}{g(v_2)-g(v_1)}
\end{equation*}
and~$\tau = \sqrt{T}$. This equation has roots
\begin{gather}
    \label{eq:tPhTr1}
    \tau_1 = \frac{m \sqrt{a}}{m \sqrt{a} + \sqrt{h}}, \\
    \label{eq:tPhTr2}
    \tau_2 = \frac{m \sqrt{a}}{m \sqrt{a} - \sqrt{h}}.
\end{gather}
It is worth noting that
\begin{equation*}
    \lim\limits_{v_2 \rightarrow v_1} h = \frac{R \big( v_1 (v_1-2d) - b c + b d + c d \big) \big( v1^2 + 2 b v_1 - b^2 \big)^2}{2 (v_1 + b) (v_1 - b)^2 (v_1 - c)^2}.
\end{equation*}

Mark that the equivalent condition to expressions~\ref{eq:PhTr1}, \ref{eq:PhTr2} is the Maxwell's rule
\begin{equation*}
    \int\limits_{v_1}^{v_2} (p(v,T) - p_0) dv = 0,
\end{equation*}
where~$p_0$ is the pressure at the phase transition and $p(v,T)$ is the thermic equation of state.

\section{Gases}
\label{sec:sim}

We use the dimensionless MSLV EoS for computations~\cite{bib:Mo_2022}
\begin{equation}
    \label{eq:EoS1_Dimless}
    p_r = \frac{T_r}{Z (v_r-b_r)} \frac{v_r-d_r}{v_r-c_r} - \frac{q_r (T_r)}{Z^2 (v_r^2 + 2 b_r v_r - b_r^2)},
\end{equation}
where~$p_r = \frac{p}{p_c}$, $v_r = \frac{v}{v_c}$, $T_r = \frac{T}{T_c}$, $b_r = \frac{b}{v_c}$, $d_r = \frac{d}{v_c}$, $c_r = \frac{c}{v_c}$, $q_r(T_r) = a_r \alpha(T_r)$, $a_r = \frac{p_c a}{(R T)^2}$, $Z = \frac{p_c v_c}{R T_c}$ and $p_c$ is the critical pressure, $v_c$ is the critical molar volume.

The constants of the dimensionless equation of state~\eqref{eq:EoS1_Dimless} for methane:
\begin{gather*}
    a_r = 0.4902264,\quad
    b_r = 0.2989634,\quad
    c_r = 0.3603434,\quad
    d_r = 0.3604034,\\
    \omega = 0.011,\quad
    m = 0.391,\quad
    Z = 0.286.
\end{gather*}
The critical values for methane:
\begin{equation*}
    p_c = 4.5992 \mathrm{MPa},\quad
    v_c = 98.63 \mathrm{cm^3/mol},\quad
    T_c = 190.56 \mathrm{K}.
\end{equation*}

We obtain that the expression~\eqref{eq:SingC2} is always positive because all constants are positive and the functions~$q''(T)$ and $G(v)$ also are positive.

\begin{figure}[p!!!]
    \begin{minipage}[h]{0.45\linewidth}
        \centering
        \includegraphics[width=0.95\linewidth]{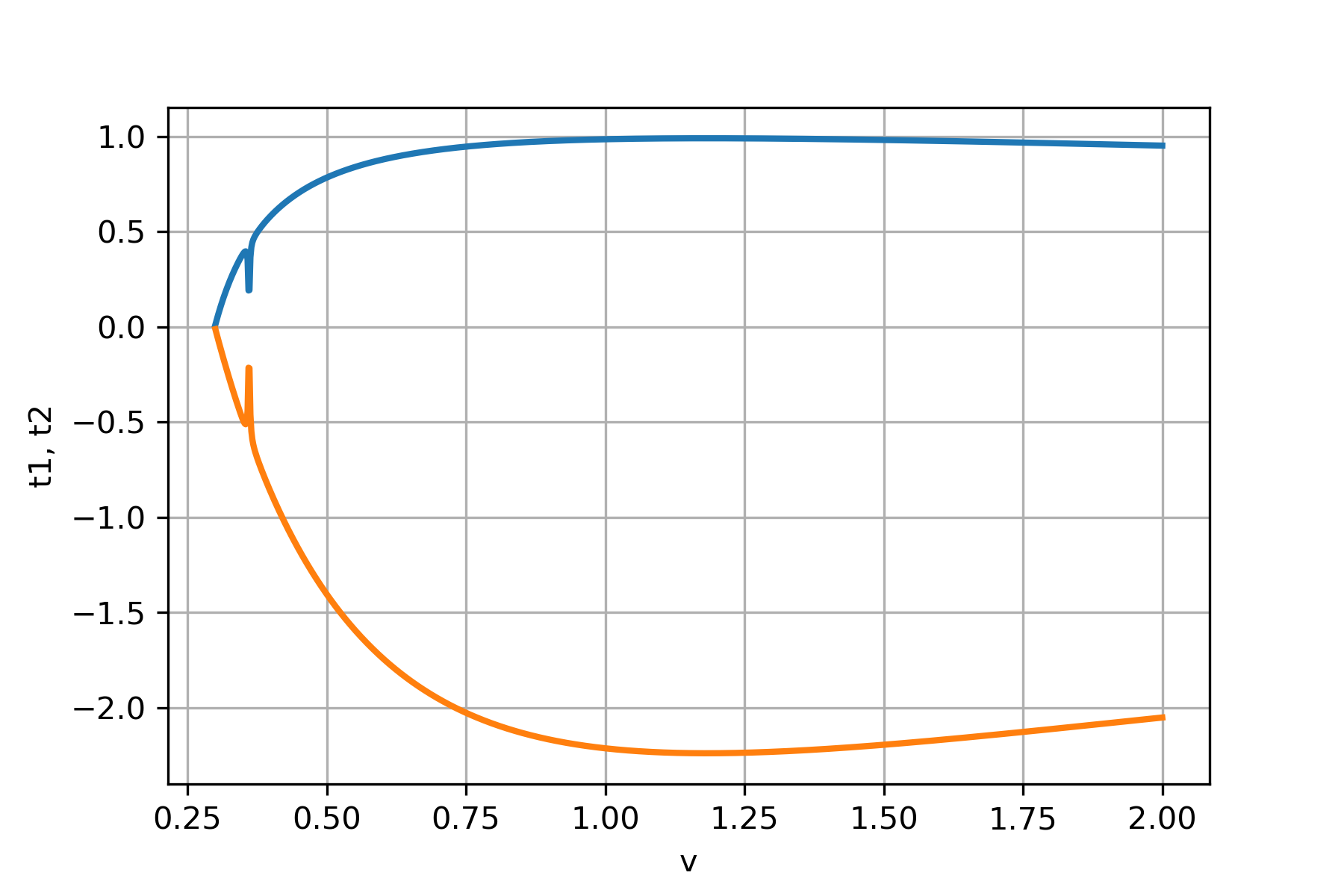}
        \caption{$\sqrt{T}$ for the methane applicable states. Blue line is~$t_1$~\eqref{eq:tphivv1}, orange -- $t_2$~\eqref{eq:tphivv2}}
        \label{fig:AS_t}
    \end{minipage}\quad
    \begin{minipage}[h]{0.45\linewidth}
        \centering
        \includegraphics[width=0.95\linewidth]{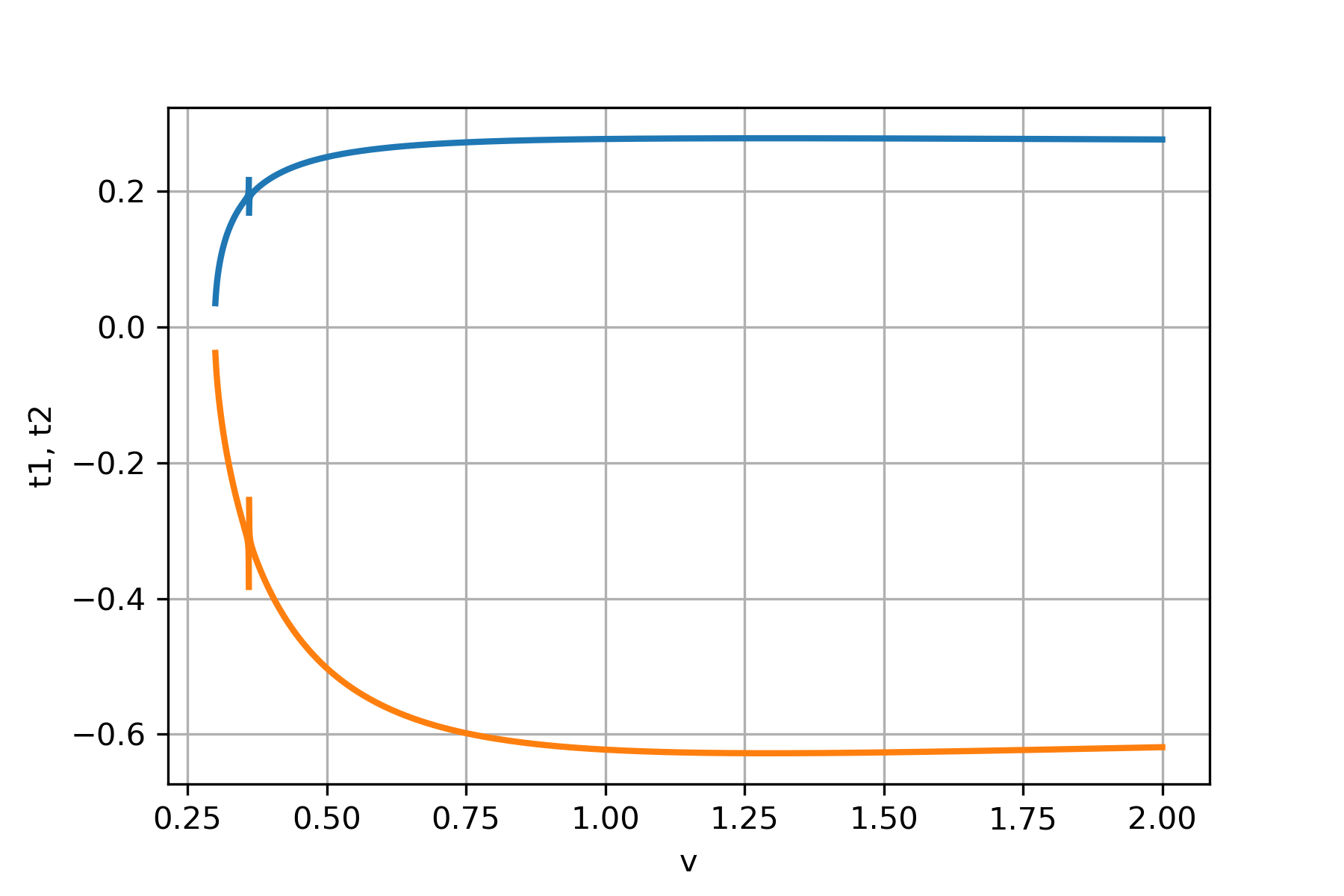}
        \caption{$\sqrt{T}$ for the methane phase transition at~$v_1 = 1$. Blue line is~$\tau_1$~\eqref{eq:tPhTr1}, orange -- $\tau_2$~\eqref{eq:tPhTr2}}
        \label{fig:PT_t}
    \end{minipage}
\end{figure}

The expressions~\eqref{eq:tphivv2} and~\eqref{eq:tPhTr2} always are negative but the expressions~\eqref{eq:tphivv1} and~\eqref{eq:tPhTr1} always are positive for methane constants of MSLV EoS. Therefore the expressions~\eqref{eq:tphivv1} and~\eqref{eq:tPhTr1} are correct for real computations. Graphics of the expressions~\eqref{eq:tphivv1}, \eqref{eq:tphivv2} and expressions~\eqref{eq:tPhTr1}, \eqref{eq:tPhTr2} for methane are given on the fig.~\ref{fig:AS_t} and~\ref{fig:PT_t} respectively. The discontinuity of functions is drawn as a ''splash''.

\begin{figure}[p!!!]
    \begin{minipage}[h]{0.45\linewidth}
        \centering
        \includegraphics[width=0.95\linewidth]{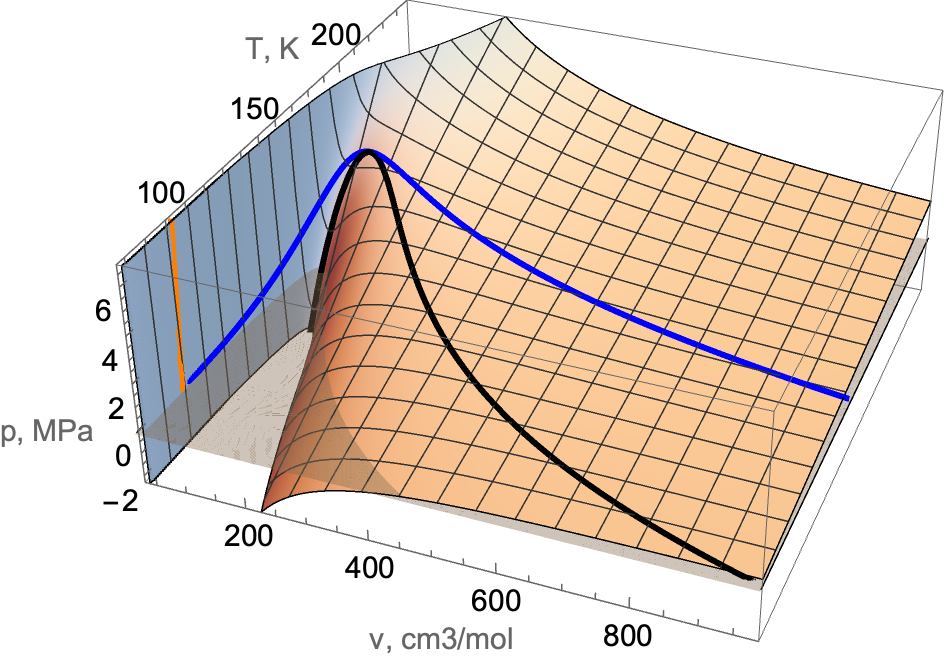}
        \caption{Lagrange manifold of methane in~$(v,T,p)$ at~$v>c$. There is no solid phase on this plot because the solid phase exists at~$v<d<c$. Black curve is the singularity curve, blue curve~-- the liquid-vapor phase transition curve, orange curve~-- one branch of the solid-liquid phase transition curve}
        \label{fig:LM1_m}
    \end{minipage}\quad
    \begin{minipage}[h]{0.45\linewidth}
        \centering
        \includegraphics[width=0.95\linewidth]{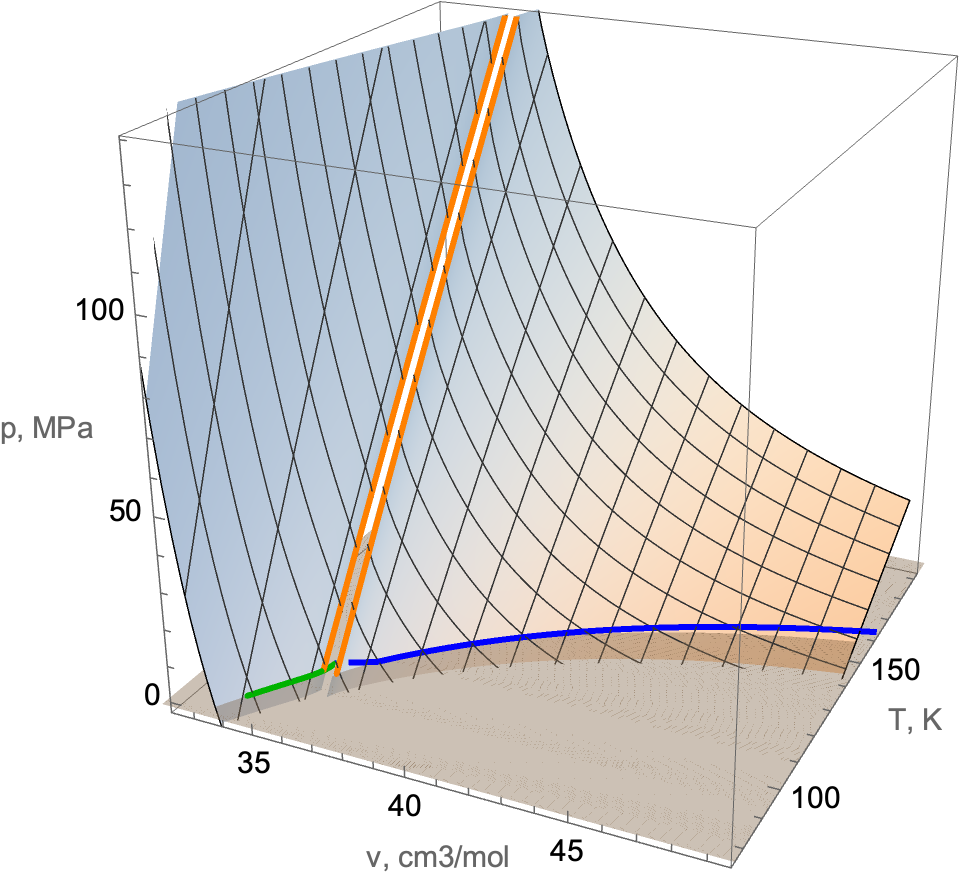}
        \caption{Lagrange manifold of methane in~$(v,T,p)$ at~$v>b$. Green line is one branch of the solid-vapor phase transition curve}
        \label{fig:LM2_m}
    \end{minipage}
\end{figure}

\begin{figure}[p!!!]
    \centering
    \includegraphics[width=100mm]{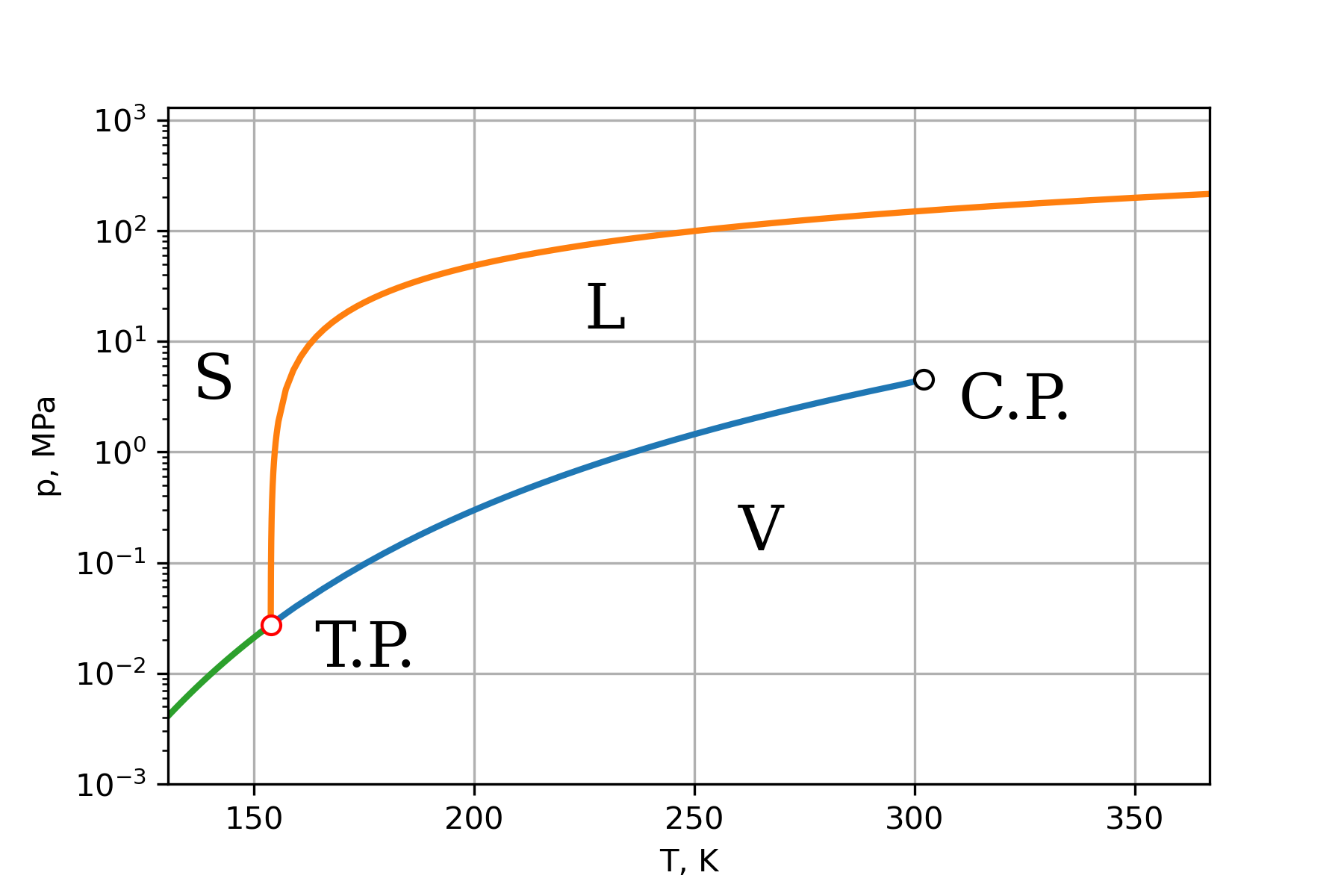}
    \caption{Phase transition curve of methane in~$(T,p)$. T.P. is the triple point, C.P. is the critical point, S denotes solid phase, L denotes liquid phase, V denotes vapour phase}
    \label{fig:PT_m}
\end{figure}

The Lagrangian manifold and phase transition curves for methane in~$(v,T,p)$ coordinates are given in fig.~\ref{fig:LM1_m}, \ref{fig:LM2_m}. One list for the liquid-vapor phase of the Lagrangian manifold is given in fig.~\ref{fig:LM1_m}. There are two lists of the Lagrangian manifold in fig.~\ref{fig:LM2_m}. The continuation of the liquid-vapor list is not drawn after phase transition curve so as not to obstruct the solid list. Fig.~\ref{fig:PT_m} give phase transition curve in coordinates~$(T,p)$.

The rest of the plots are in Appendix A.

\section{Conclusion}
\label{sec:concl}

The Modified Solid-Liquid-Vapour equation of state is the extension of the Peng-Robinson EoS to describe three phases of substance. The caloric EoS and thermodynamic potentials were written out. The conditions of the applicable states and first order phase transition were considered. The PT-diagram has one critical and one triple point. The border of applicable states is the singularities of the Lagrangian manifold projection on the planes of the extensive variables and intensive variables. The three-dimensional plots with the Lagrangian manifold, the border of applicable states and phase transition curves were drawn for methane. The curves of the vapor-liquid phase transition were plotted as well as the curves of the solid-liquid and solid-vapor phase transition. Also some plots of the applicable states and phase transition projections in the plane of thermodynamic variables were drawn.

\section{Acknowledgments}

The authors express their gratitude to  Valentin V. Lychagin.

This research was partially funded by Russian Science Foundation grant number 21-71-20034.

All 2D graphics was plotted with feslib library on Python. https://github.com/LychaginTeam/feslib

The computations were performed on the MVS-10P OP supercomputer at the Joint Supercomputer Center of the Russian Academy of Sciences.

\appendix

\section*{Appendix A}

Other graphics for the phase transition and singularity curve in different planes of thermodynamic variables are presented in here.

\begin{figure}[p!!!]
    \begin{minipage}[h]{0.45\linewidth}
        \centering
        \includegraphics[width=0.95\linewidth]{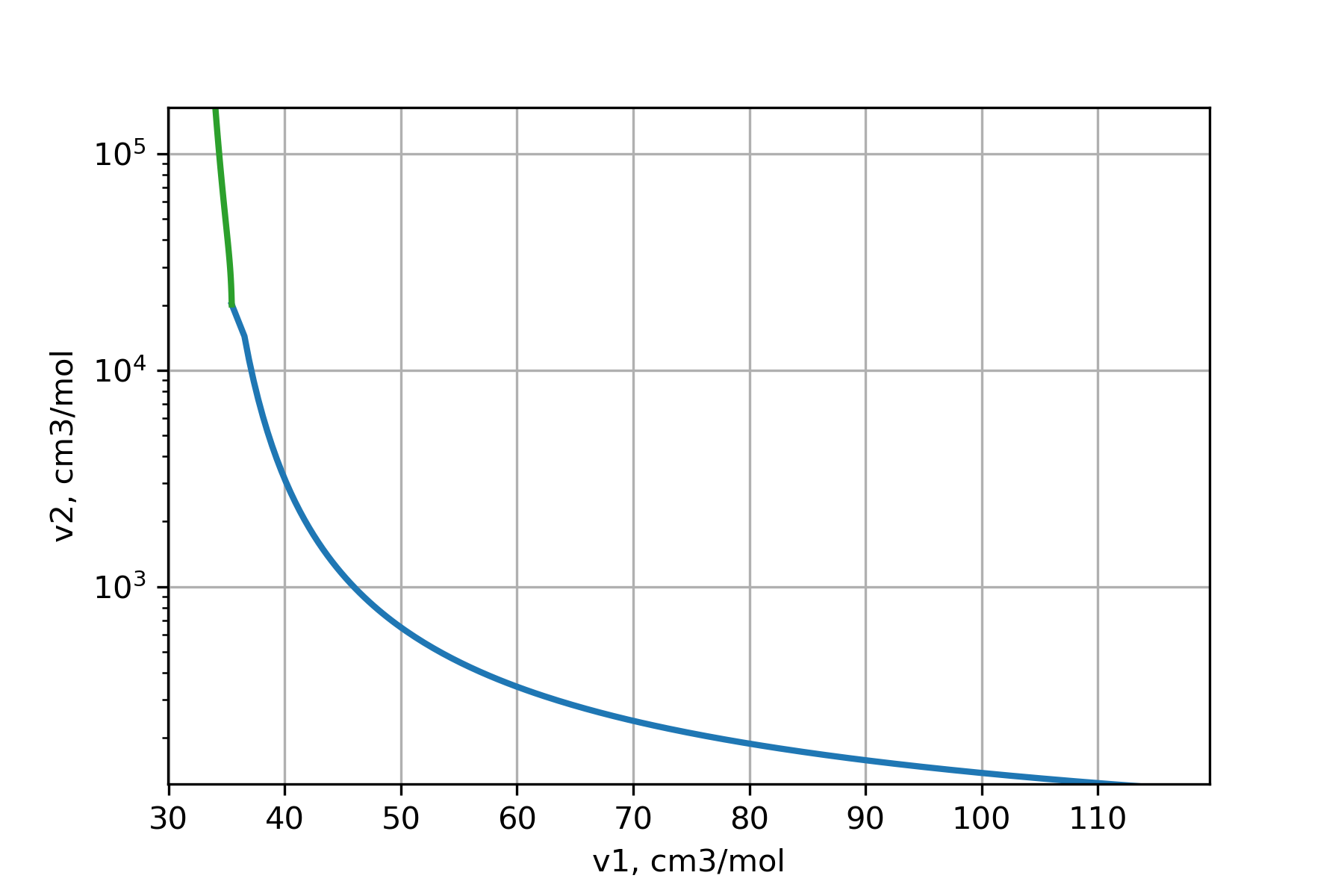}
        \caption{S-G, L-G phase transition points for methane in plane~$(v_1,v_2)$. Logarithmic scale for~$v_2$}
    \end{minipage}\quad
    \begin{minipage}[h]{0.45\linewidth}
        \centering
        \includegraphics[width=0.95\linewidth]{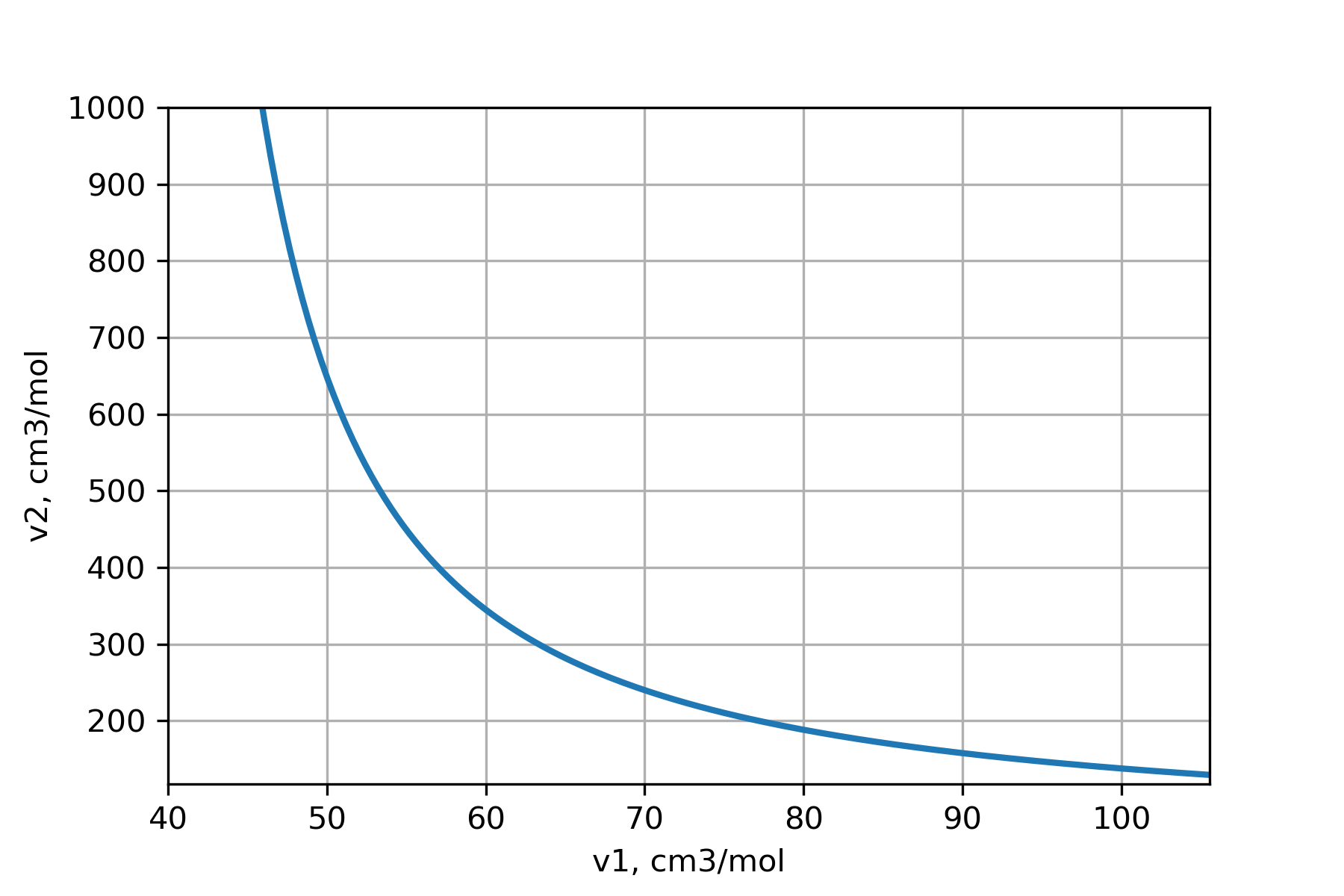}
        \caption{L-G phase transition points for methane in plane~$(v_1,v_2)$}
    \end{minipage}
\end{figure}

\begin{figure}[p!!!]
    \begin{minipage}[h]{0.45\linewidth}
        \centering
        \includegraphics[width=0.95\linewidth]{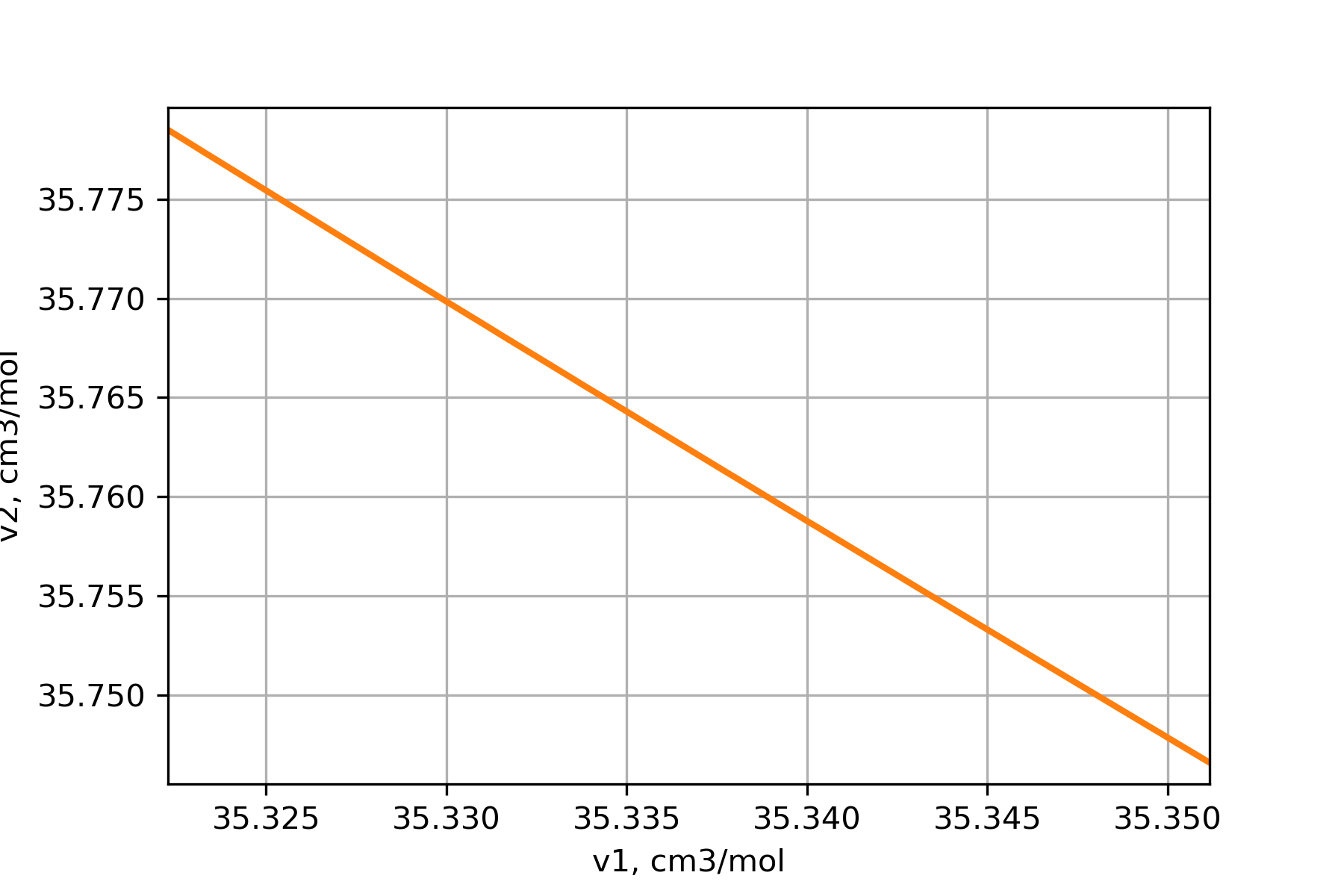}
        \caption{S-L phase transition points for methane in plane~$(v_1,v_2)$}
    \end{minipage}\quad
    \begin{minipage}[h]{0.45\linewidth}
        \centering
        \includegraphics[width=0.95\linewidth]{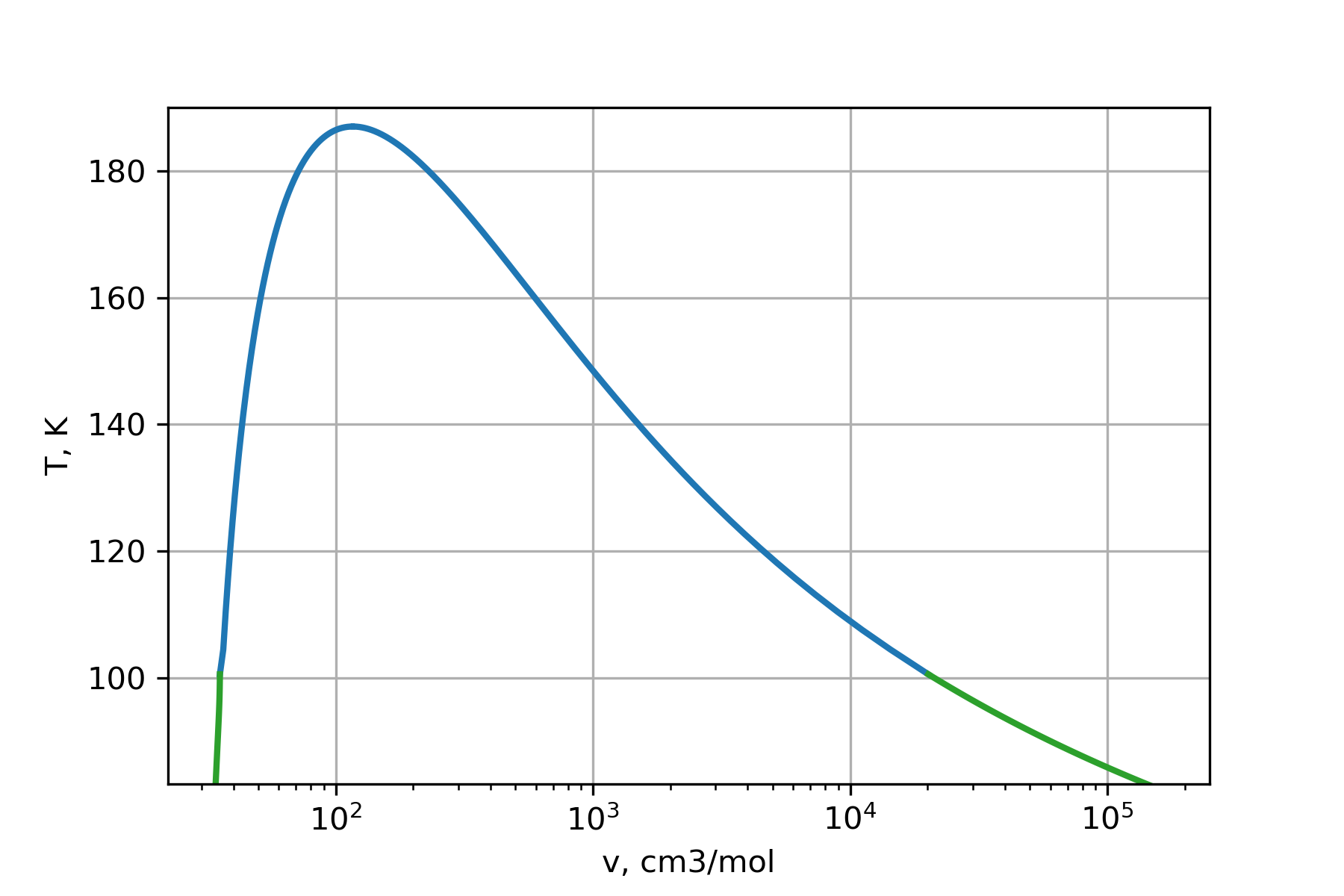}
        \caption{S-G, L-G phase transition points for methane in plane~$(v,T)$. Logarithmic scale for~$v$}
    \end{minipage}
\end{figure}

\begin{figure}[p!!!]
    \begin{minipage}[h]{0.45\linewidth}
        \centering
        \includegraphics[width=0.95\linewidth]{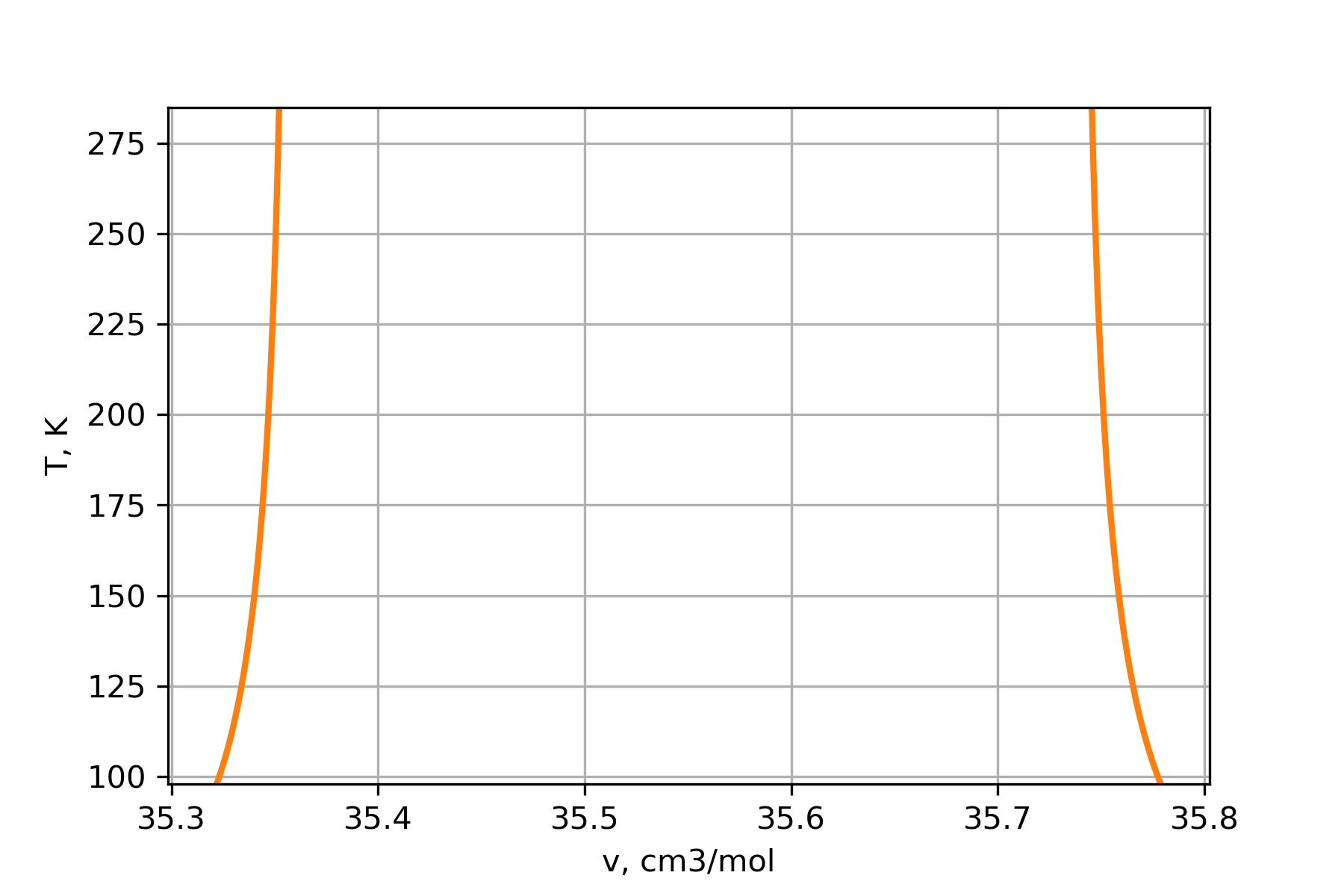}
        \caption{S-L phase transition points for methane in plane~$(v,T)$}
    \end{minipage}\quad
    \begin{minipage}[h]{0.45\linewidth}
        \centering
        \includegraphics[width=0.95\linewidth]{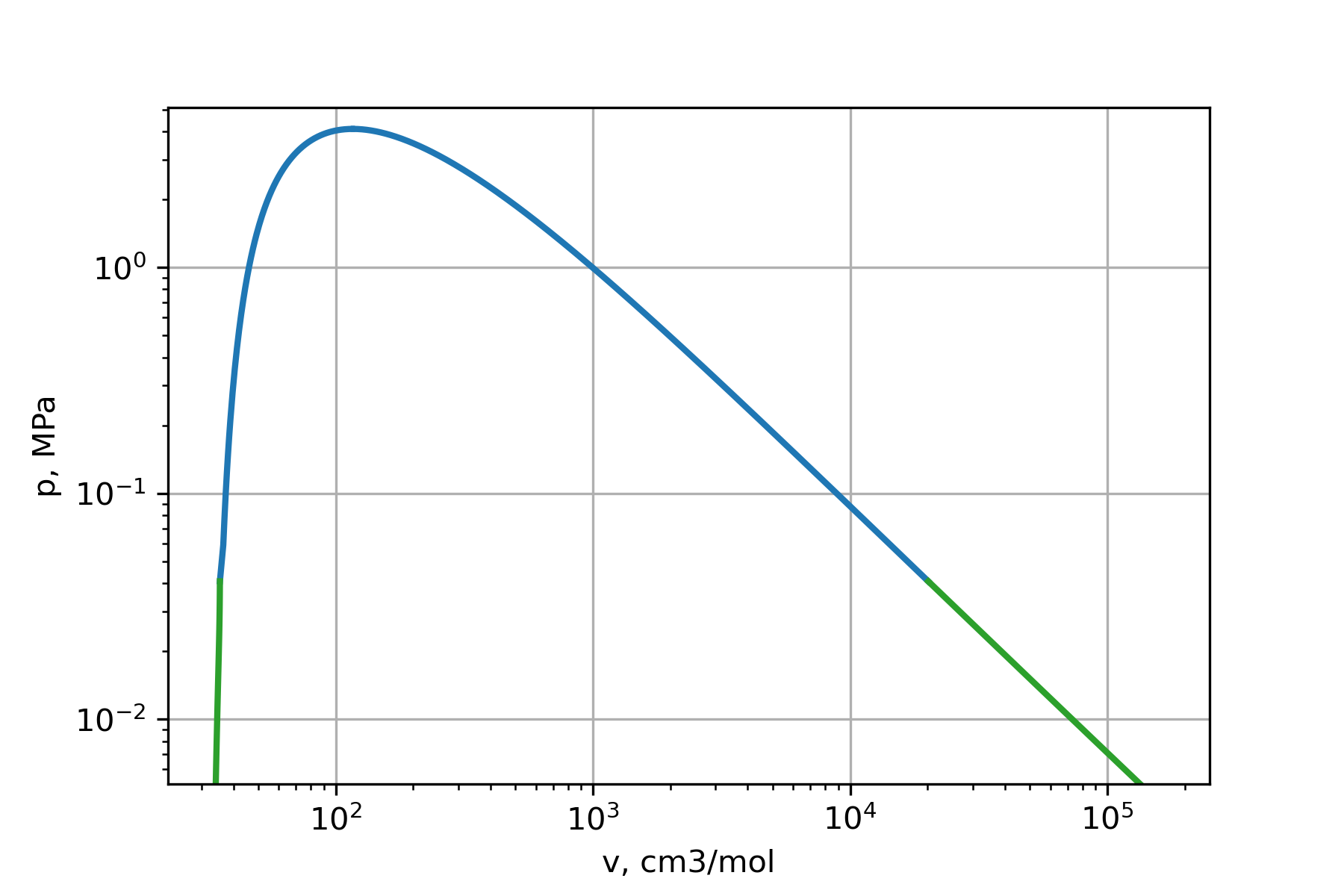}
        \caption{S-G, L-G phase transition points for methane in plane~$(v,p)$. Logarithmic scale for~$v$ and $p$}
    \end{minipage}
\end{figure}

\begin{figure}[p!!!]
    \begin{minipage}[h]{0.45\linewidth}
        \centering
        \includegraphics[width=0.95\linewidth]{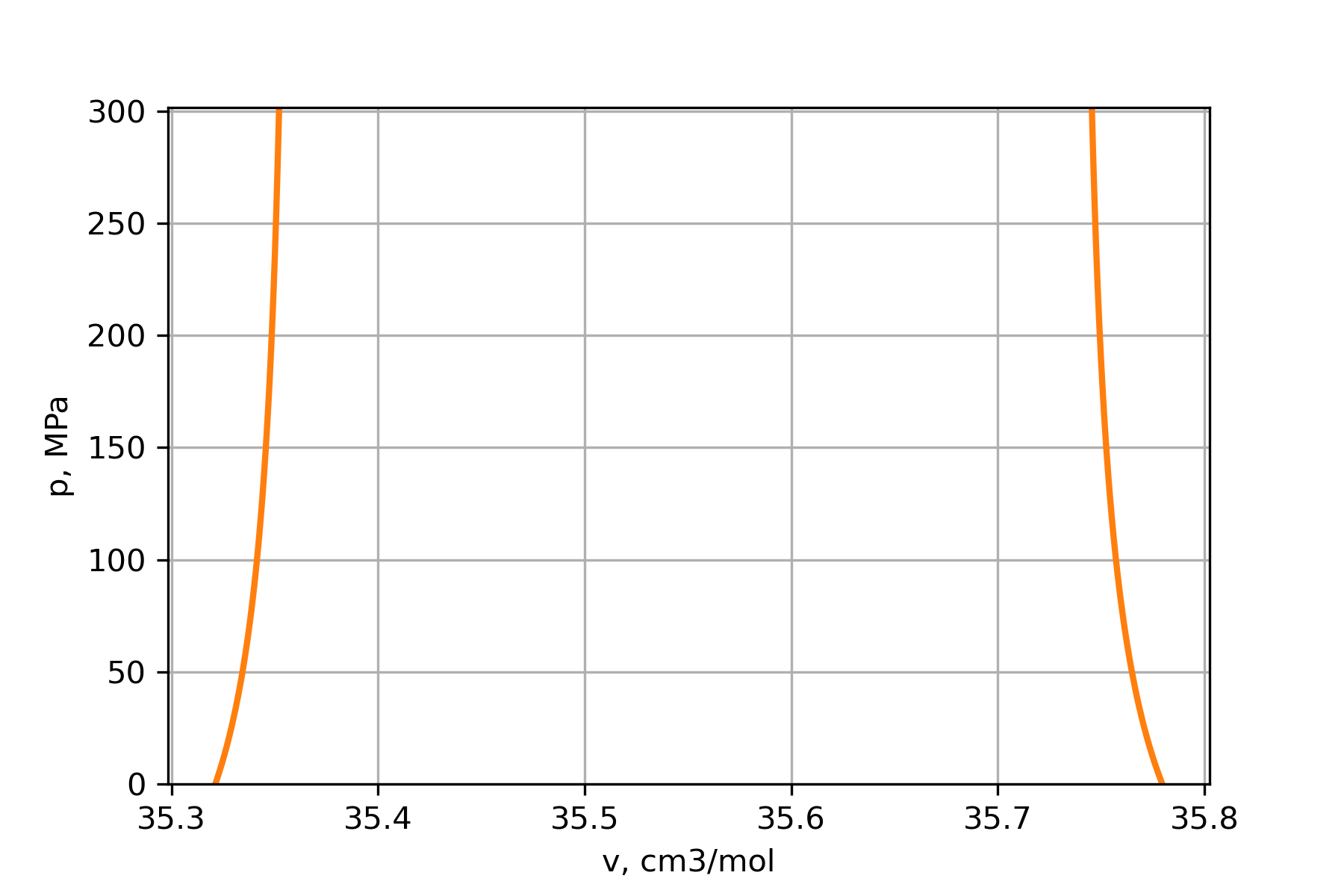}
        \caption{S-L phase transition points for methane in plane~$(v,p)$}
    \end{minipage}\quad
    \begin{minipage}[h]{0.45\linewidth}
        \centering
        \includegraphics[width=0.95\linewidth]{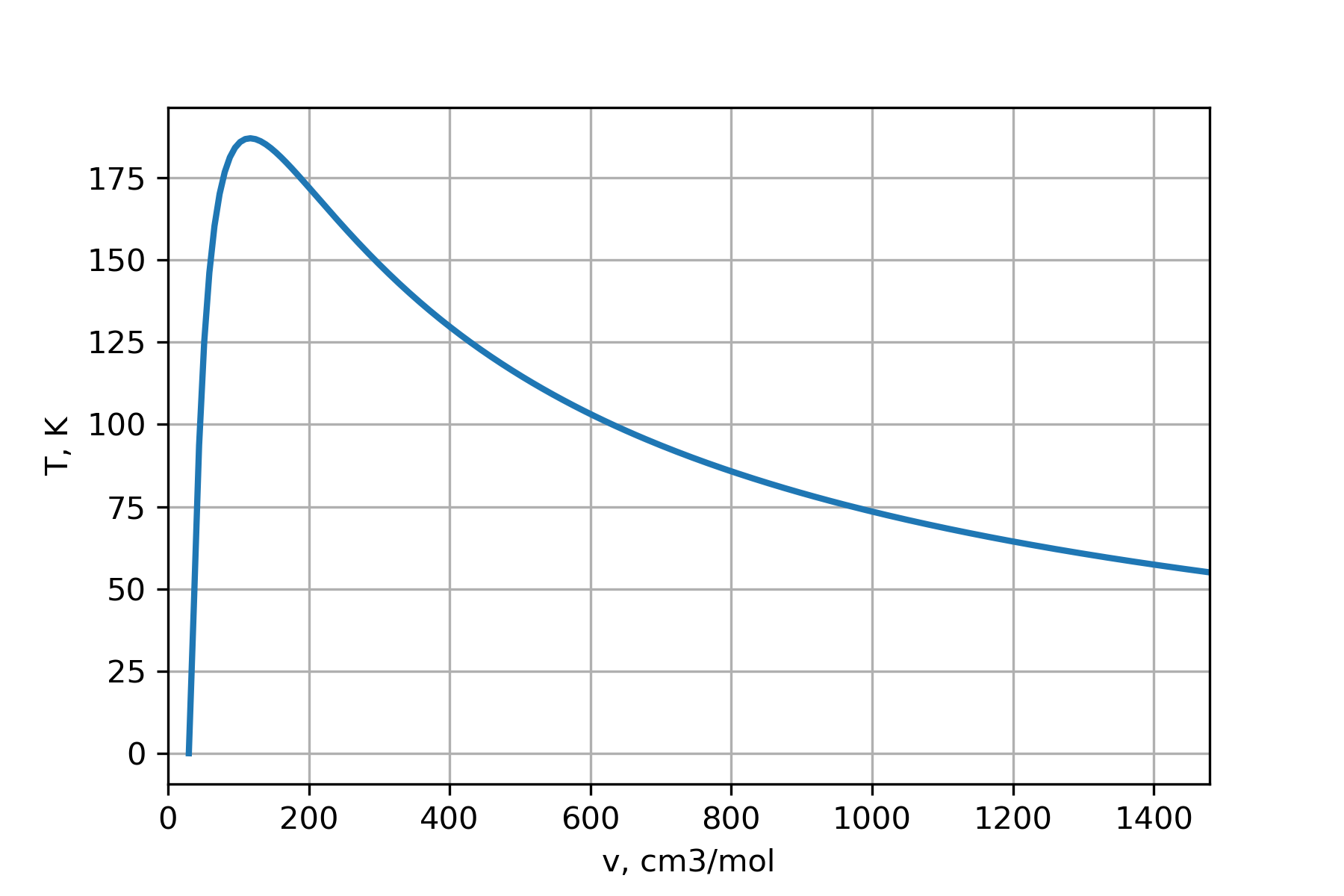}
        \caption{Singularity curve for methane in plane~$(v,T)$}
    \end{minipage}
\end{figure}

\begin{figure}[p!!!]
    \begin{minipage}[h]{0.45\linewidth}
        \centering
        \includegraphics[width=0.95\linewidth]{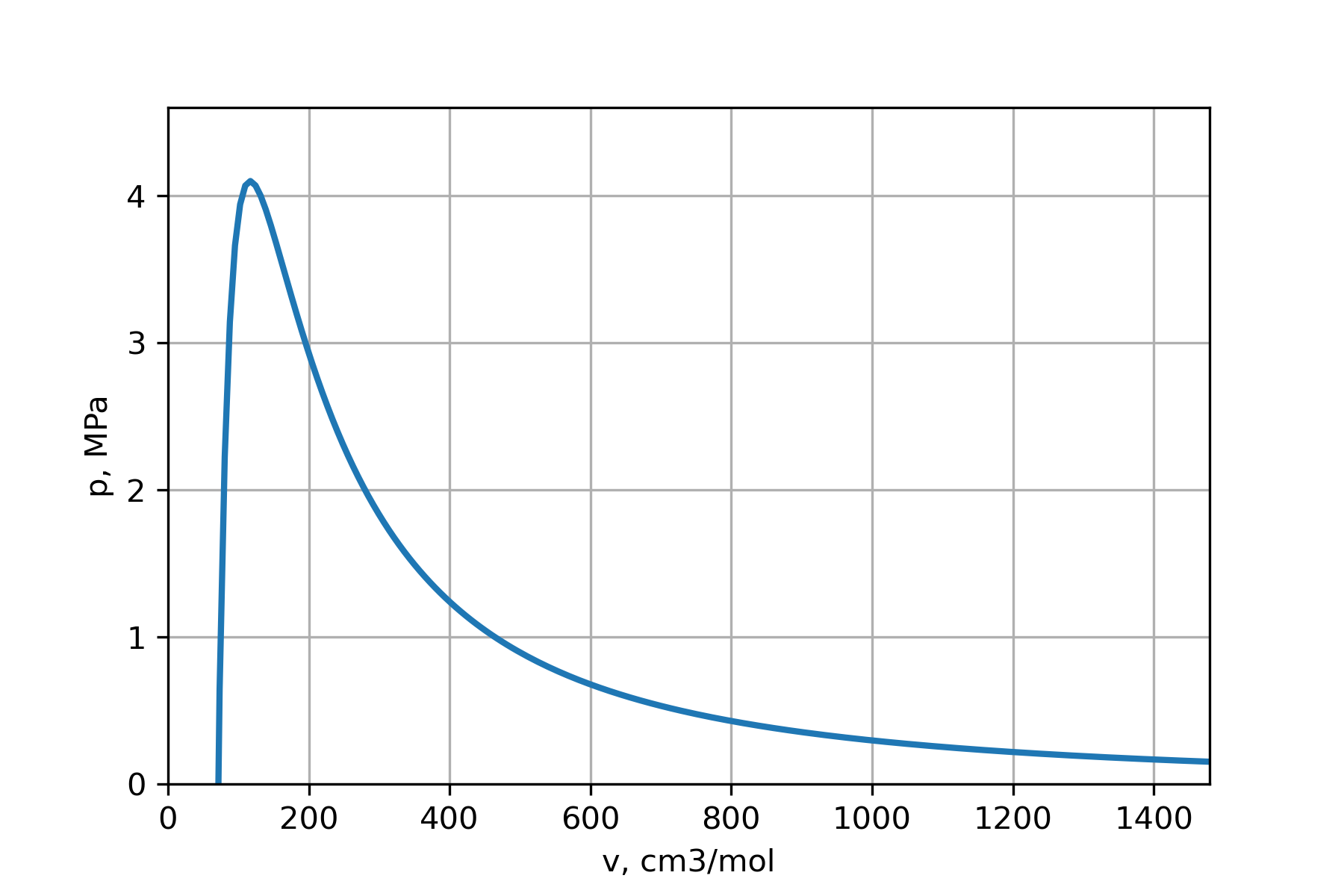}
        \caption{Singularity curve for methane in plane~$(v,p)$}
    \end{minipage}\quad
    \begin{minipage}[h]{0.45\linewidth}
        \centering
        \includegraphics[width=0.95\linewidth]{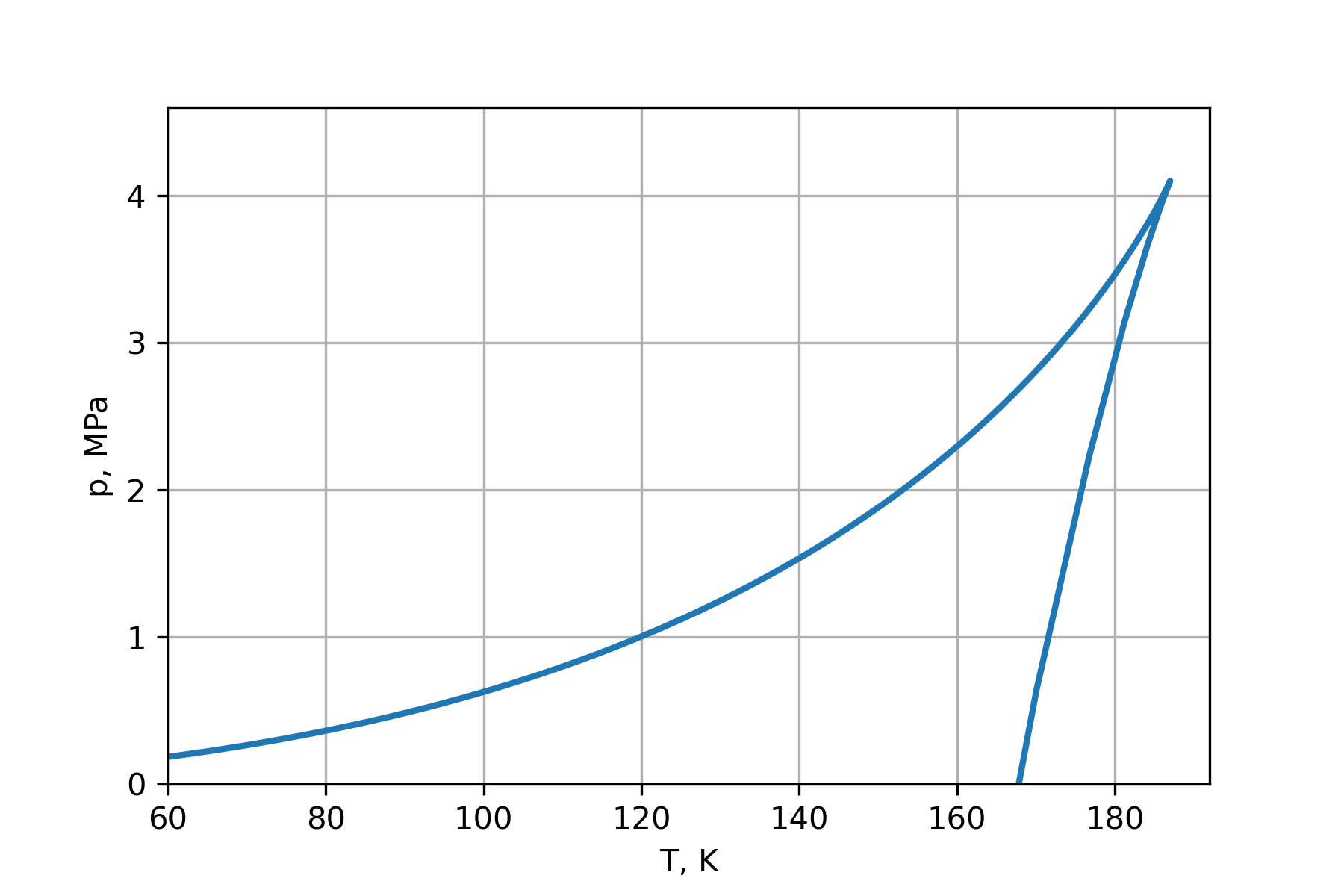}
        \caption{Singularity curve for methane in plane~$(T,p)$}
    \end{minipage}
\end{figure}

\end{document}